\documentstyle[12pt,amssymbols]{article}

\def\journal{\topmargin .3in	\oddsidemargin .5in
	\headheight 0pt	\headsep 0pt
	\textwidth 5.625in 
	\textheight 8.25in 
	\marginparwidth 1.5in
	\parindent 2em
	\parskip .5ex plus .1ex		\jot = 1.5ex}
\journal

\catcode`\@=11
\def\marginnote#1{}
\newcount\hour
\newcount\minute
\newtoks\amorpm
\hour=\time\divide\hour by60
\minute=\time{\multiply\hour by60 \global\advance\minute by-\hour}
\edef\standardtime{{\ifnum\hour<12 \global\amorpm={am}%
	\else\global\amorpm={pm}\advance\hour by-12 \fi
	\ifnum\hour=0 \hour=12 \fi
	\number\hour:\ifnum\minute<10 0\fi\number\minute\the\amorpm}}
\edef\militarytime{\number\hour:\ifnum\minute<10 0\fi\number\minute}
\def\draftlabel#1{{\@bsphack\if@filesw {\let\thepage\relax
   \xdef\@gtempa{\write\@auxout{\string
      \newlabel{#1}{{\@currentlabel}{\thepage}}}}}\@gtempa
   \if@nobreak \ifvmode\nobreak\fi\fi\fi\@esphack}
	\gdef\@eqnlabel{#1}}
\def\@eqnlabel{}
\def\@vacuum{}
\def\draftmarginnote#1{\marginpar{\raggedright\scriptsize\tt#1}}
\def\draft{\oddsidemargin -.5truein
	\def\@oddfoot{\sl preliminary draft \hfil
	\rm\thepage\hfil\sl\today\quad\militarytime}
	\let\@evenfoot\@oddfoot	\overfullrule 3pt
	\let\label=\draftlabel
	\let\marginnote=\draftmarginnote
   \def\@eqnnum{(\theequation)\rlap{\kern\marginparsep\tt\@eqnlabel}%
\global\let\@eqnlabel\@vacuum}  }
\def\preprint{\twocolumn\sloppy\flushbottom\parindent 2em
	\leftmargini 2em\leftmarginv .5em\leftmarginvi .5em
	\oddsidemargin -.5in	\evensidemargin -.5in
	\columnsep .4in	\footheight 0pt
	\textwidth 10in	\topmargin  -.4in
	\headheight 12pt \topskip .4in
	\textheight 7.1in \footskip 0pt
	\def\@oddhead{\thepage\hfil\addtocounter{page}{1}\thepage}
	\let\@evenhead\@oddhead	\def\@oddfoot{}	\def\@evenfoot{} }
\def\numberbysection{\@addtoreset{equation}{section}
	\def\theequation{\thesection.\arabic{equation}}}
\def\underline#1{\relax\ifmmode\@@underline#1\else
	$\@@underline{\hbox{#1}}$\relax\fi}
\def\titlepage{\@restonecolfalse\if@twocolumn\@restonecoltrue\onecolumn
     \else \newpage \fi \thispagestyle{empty}\c@page\z@
	\def\thefootnote{\fnsymbol{footnote}} }
\def\endtitlepage{\if@restonecol\twocolumn \else \newpage \fi
	\def\thefootnote{\arabic{footnote}}
	\setcounter{footnote}{0}}  
\catcode`@=12
\relax
\def\figcap{\section*{Figure Captions\markboth
	{FIGURECAPTIONS}{FIGURECAPTIONS}}\list
	{Figure \arabic{enumi}:\hfill}{\settowidth\labelwidth
{Figure 999:}
	\leftmargin\labelwidth
	\advance\leftmargin\labelsep\usecounter{enumi}}}
 \relax
\def\tablecap{\section*{Table Captions\markboth
	{TABLECAPTIONS}{TABLECAPTIONS}}\list
	{Table \arabic{enumi}:\hfill}{\settowidth\labelwidth{Table 999:}
	\leftmargin\labelwidth
	\advance\leftmargin\labelsep\usecounter{enumi}}}
 \relax
\def\reflist{\section*{References\markboth
	{REFLIST}{REFLIST}}\list
	{[\arabic{enumi}]\hfill}{\settowidth\labelwidth{[999]}
	\leftmargin\labelwidth
	\advance\leftmargin\labelsep\usecounter{enumi}}}
 \relax
\makeatletter
\newcounter{pubctr}
\def\publist{\@ifnextchar[{\@publist}{\@@publist}}
\def\@publist[#1]{\list
	{[\arabic{pubctr}]\hfill}{\settowidth\labelwidth{[999]}
	\leftmargin\labelwidth
	\advance\leftmargin\labelsep
	\@nmbrlisttrue\def\@listctr{pubctr}
	\setcounter{pubctr}{#1}\addtocounter{pubctr}{-1}}}
\def\@@publist{\list
	{[\arabic{pubctr}]\hfill}{\settowidth\labelwidth{[999]}
	\leftmargin\labelwidth
	\advance\leftmargin\labelsep
	\@nmbrlisttrue\def\@listctr{pubctr}}}
 \relax
\makeatother
\catcode`\@=11
\def\section{\@startsection {section}{1}{0pt}{-3.5ex plus -1ex minus
 -.2ex}{2.3ex plus .2ex}{\raggedright\large\bf}}
\catcode`\@=12
\newskip\humongous \humongous=0pt plus 1000pt minus 1000pt

\newif\ifdtup

\def\oldreffmt#1{\rlap{[#1]} \hbox to 2\parindent{}}

\def\figfmt#1{\rlap{Figure {#1}} \hbox to 1in{}}

\def\etal{\hbox{\it et al.}}

\def\abs#1{\left| #1\right|}

\def\beq{\begin{equation}}
\def\eeq{\end{equation}}

\def\bea{\begin{eqnarray}}

\def\eea{\end{eqnarray}}
\catcode`\@=11
\def\eqnarray{\stepcounter{equation}\let\@currentlabel=\theequation
\global\@eqnswtrue
\global\@eqcnt\z@\tabskip\@centering\let\\=\@eqncr
\gdef\@@fix{}\def\eqno##1{\gdef\@@fix{##1}}%
$$\halign to \displaywidth\bgroup\@eqnsel\hskip\@centering
  $\displaystyle\tabskip\z@{##}$&\global\@eqcnt\@ne
  \hskip 2\arraycolsep \hfil${##}$\hfil
  &\global\@eqcnt\tw@ \hskip 2\arraycolsep
$\displaystyle\tabskip\z@{##}$\hfil
   \tabskip\@centering&\llap{##}\tabskip\z@\cr}

\def\@@eqncr{\let\@tempa\relax
    \ifcase\@eqcnt \def\@tempa{& & &}\or \def\@tempa{& &}
      \else \def\@tempa{&}\fi
     \@tempa \if@eqnsw\@eqnnum
\stepcounter{equation}\else\@@fix\gdef\@@fix{}\fi
     \global\@eqnswtrue\global\@eqcnt\z@\cr}

\catcode`\@=12
\font\tenbifull=cmmib10 
\font\tenbimed=cmmib10 scaled 800
\font\tenbismall=cmmib10 scaled 666
\relax

\def\Rp{$\not \!\! R_p\,$}
\def\np#1#2#3{        {Nucl. Phys. }{\bf #1}, #2 (19#3)}
\def\pl#1#2#3{        {Phys. Lett. }{\bf #1}, #2 (19#3)}
\def\pr#1#2#3{        {Phys. Rev. }{\bf #1}, #2 (19#3)}
\def\prpt#1#2#3{      {Phys. Rep. }{\bf #1}, #2 (19#3)}
\def\prgth#1#2#3{      {Prog. Theor. Phys. }{\bf#1}, #2 (19#3)}
\def\prl#1#2#3{       {Phys. Rev. Lett. }{\bf #1}, #2 (19#3)}
\def\arn#1#2#3{       {Ann. Rev. Nucl. Part. Sci. }{\bf #1}, #2 (19#3)}
\def\mpl#1#2#3{           {Mod. Phys. Lett. }{\bf #1}, #2 (19#3)}
\def\npp#1#2#3{       {Nucl. Phys. }{\bf (Proc. Suppl.) #1}, #2 (19#3)}

\def\thefootnote{\fnsymbol{footnote}}
\begin{document}
\begin{titlepage}
\today          \hfill
\begin{center}
\hfill    LBL-37823 \\
          \hfill    UCB-PTH-95/33 \\

\vskip .5in

{\large \bf $R$-Parity Violation in Flavor Changing Neutral Current
Processes
and Top Quark Decays}
\footnote{This work was supported in part by the Director, Office of
Energy Research, Office of High Energy and Nuclear Physics, Division of
High Energy Physics of the U.S. Department of Energy under Contract
DE-AC03-76SF00098 and in part by the National Science Foundation under
grant PHY-90-21139.}

\vskip .5in

K. Agashe\footnote{email: agashe@theor1.lbl.gov} and
M. Graesser\footnote
{email: graesser@theor1.lbl.gov} \footnote{Supported
by a Natural
Sciences and Engineering Research Council of Canada Fellowship.}
\\

{\em Theoretical Physics Group\\
    Lawrence Berkeley National Laboratory\\
      University of California\\
    Berkeley, California 94720}
\end{center}

\vskip .225in
\begin{abstract}

We show that supersymmetric $R$-parity breaking ($\not \! \! R_p$)
interactions always result in Flavor Changing Neutral Current
(FCNC) processes.
Within a single coupling scheme, these processes can be avoided in
either the charge $+2/3$ or the charge $-1/3$ quark sector,
but not both.
These processes are used to place constraints on \Rp couplings.
The constraints on the first and the second generations are better than
those existing in the literature.
The \Rp interactions may result in new top quark decays. Some of these
violate electron-muon universality or produce a
 surplus of $b$ quark events in $t\bar{t}$ decays. Results from the CDF
experiment are used to bound these \Rp
couplings.
\end{abstract}
\end{titlepage}
\renewcommand{\thepage}{\roman{page}}
\setcounter{page}{2}

\vskip 1in

\begin{center}
{\bf Disclaimer}
\end{center}

\vskip .2in

\begin{scriptsize}
\begin{quotation}

This document was prepared as an account of work sponsored by the United
States Government. While this document is believed to contain correct
 information, neither the United States Government nor any agency
thereof, nor The Regents of the University of California, nor any of their
employees, makes any warranty, express or implied, or assumes any legal
liability or responsibility for the accuracy, completeness, or usefulness
of any information, apparatus, product, or process disclosed, or represents
that its use would not infringe privately owned rights.  Reference herein
to any specific commercial products process, or service by its trade name,
trademark, manufacturer, or otherwise, does not necessarily constitute or
imply its endorsement, recommendation, or favoring by the United States
Government or any agency thereof, or The Regents of the University of
California.  The views and opinions of authors expressed herein do not
necessarily state or reflect those of the United States Government or any
agency thereof, or The Regents of the University of California.
\end{quotation}
\end{scriptsize}

\vskip 2in

\begin{center}
\begin{small}
{\it Lawrence Berkeley Laboratory is an equal opportunity employer.}
\end{small}
\end{center}

\newpage
\renewcommand{\thepage}{\arabic{page}}
\setcounter{page}{1}

\section{Introduction}

The Minimal Supersymmetric Standard Model (MSSM) \cite{nilles} with
the gauge group $G=SU(3)_c\times SU(2)_L\times U(1)_Y$ contains
the Standard Model
particles and their superpartners, and an additional Higgs doublet.
In order to produce the observed spectrum of particle masses,
the superpotential is given by
\begin{equation}
\lambda^L_{ij}L_iE_j^cH + \lambda^D_{jk}HQ_jD^c_k +
\lambda^U_{ij}U^c_iQ_jH^{\prime} + \mu HH^{\prime}
\end{equation}
where $L=\left( \begin{array}{c} N  \\ E\\
\end{array} \right)$ and $Q=\left( \begin{array}{c} U \\ D\\
\end{array} \right)$ denote the chiral superfields containing the
lepton and quark $SU(2)_L$ doublets
and $E^c$, $U^c$ and $D^c$ are the $SU(2)_L$ singlets, all in the weak
basis. $H$ and $H^{\prime}$ are the Higgs
doublets
with hypercharges $-1$ and $+1$ respectively.
The $SU(2)_L$ and $SU(3)_c$ indices are suppressed, and $i,j$ and $k$ are
generation indices. However, requiring the Lagrangian to
be gauge invariant does not uniquely determine the form of the
superpotential. In addition, the following renormalizable
 terms
\begin{equation}
\lambda_{ijk}L_iL_jE_k^c + \bar{\lambda}_{ijk}L_iQ_jD^c_k +
\lambda^{\prime\prime}_{ijk}U^c_iD^c_jD^c_k
\end{equation}
are allowed\footnote{A term $\mu_iL_iH^{\prime}$ is also allowed.
This may be rotated away through a redefinition of the $L$ and
$H$
fields \cite{suzuki}.}. Unlike the interactions of the MSSM, these
terms violate lepton number and baryon number.
They may be forbidden by imposing a discrete symmetry, $R$-parity,
which is ${(-1)}^{3B+L+2S}$ on a component field with baryon
number $B$, lepton number $L$ and spin $S$.
Whether this symmetry is realized in nature must be determined by
experiment.
If both lepton and baryon number violating
interactions are present, then limits on the proton lifetime place
stringent constraints on the products of most of
these couplings.
So,
it is usually assumed that if $R$-parity is violated, then either
lepton or baryon number violating interactions, but
not both, are
present. It is interesting that despite the large limits on the
proton lifetime, some products of the $R$-parity violating
couplings remain
bounded only by the requirement
that the theory remain perturbative \cite{carlson}. If either
$L_iQ_jD_k^c$ or $U^c_iD^c_jD^c_k$ terms are present, flavor
changing neutral current (FCNC) processes are induced. It has been
assumed that if only one $R$-parity violating (\Rp)
coupling with a particular
flavor structure is non-zero, then these flavor changing processes
are avoided. In this \it single coupling scheme \rm \cite{hall}
 then,
efforts at constraining $R$-parity violation have concentrated on
flavor conserving processes
\cite{barger,godbole,dawson,choudhury,ellis,mohapatra}.
It is surprising that, even though
individual lepton or baryon number is violated in this scheme,
the constraints are rather weak.

In Section 2, we demonstrate that the \it single coupling scheme
\rm cannot be realized in the quark mass basis.
Despite the general values the couplings may have in the weak basis,
after electroweak symmetry
breaking there is at least one large \Rp coupling and many other \Rp
couplings with different flavor
structure. Therefore, in the mass basis the $R$-parity
breaking couplings \it cannot \rm be diagonal in generation space.
Thus, flavor changing neutral current processes are always
present
in either the charge $2/3$ or the charge $-1/3$ quark sectors.
We use these processes to place constraints on $R$-parity breaking.
We find constraints on the first and the second generations that are
much stronger than existing limits.

The recent discovery of the top quark \cite{D0,CDF1} with the large
mass of $176$ $GeV$ opens the possibility for
the tree level decays $t\rightarrow
\tilde{l}^+_i + d_k$ and $t\rightarrow \bar{\tilde{d}_j} + \bar{d_k}$
if $R$-parity is broken. If the \Rp
couplings are large enough, then these decay channels may be
competitive with the Standard Model decay $t\rightarrow b+W$. As no
inconsistencies between the measured branching fractions and
production cross-section of the top quark and those
predicted by the Standard Model (SM) have been reported, limits on
the branching fractions for the $\not\!\!R_p$ decay channels may
be obtained. Since the existing lower bound on the mass of the
lightest slepton is $\sim 45$ $GeV$ \cite{pdg}, while the strong
interactions of the squarks make it
likely that the squarks are heavier than the sleptons, the decay
$t\rightarrow \tilde{l}^+_i + d_k$ is more probable. In our
analysis, we therefore assume that only the slepton decay channel is
present. In Section 3 we analyse the $\not\!\!R_p$ top decay
channels to place constraints on the $t\rightarrow \tilde{l}^+_i + d_k$
coupling. For this reason, in this paper we assume that
only the $\not\!\!L$ terms $L_iQ_jD_k^c$ are present. The conclusions of
Section 2, however, are valid even if the $L_iL_jE_k^c$ terms are also
present.
Constraints on products of couplings when both $\not\!\!L$ interactions
are
present may be found in reference \cite{ma}. In Section 4 we summarize
our
results and compare them with limits exisiting in the
literature.

\section{Flavor Changing Neutral Current Processes}
\label{sec-FCNC}

Flavor changing neutral current processes are more clearly seen by
examining the structure of the
interactions in the quark mass basis. In this basis, the
$\bar{\lambda}_{ijk}$ interactions are
\begin{equation}
\lambda^{\prime}_{ijk}(N^m_i(V_{KM})_{jl}D^m_l - E^m_iU^m_j)D^{cm}_k
\label{physbasis}
\end{equation}
where
\begin{equation}
\lambda^{\prime}_{ijk}=\bar{\lambda}_{imn}U_{Ljm}D_{Rnk}^{\ast}
\end{equation}
The superfields in Equation (\ref{physbasis}) have their fermionic
components in the mass basis so that the
Cabibbo-Kobayashi-Masakawa (CKM) matrix \cite{kobayashi}
$V_{KM}$ appears explicitly. The rotation matrices $U_L$ and $D_R$
appearing in the previous equation are defined by
\begin{eqnarray}
u_{Li}=U_{Lij}u^m_{Lj} \\
d_{Ri}=D_{Rij}d^m_{Rj}
\end{eqnarray}
where $q_i\;(q_i^m)$ are quark fields in the weak (mass) basis.
Henceforth, all the fields will be in the mass basis and we drop
the superscript $m$.

Unitarity of the rotation matrices implies that the couplings
$\lambda_{ijk}^{\prime}$ and $\bar{\lambda}_{ijk}$ satisfy
\begin{equation}
\sum_{jk} \abs{\lambda^{\prime}_{ijk}}^2 =  \sum_{mn}
\abs{\bar{\lambda}_{imn}}^2
\end{equation}
So any constraint on the \Rp couplings in the quark mass basis also
places a bound on the $\not\!R_p$
couplings
in the weak basis.

In terms of component fields, the interactions are
\begin{equation}
\lambda^{\prime}_{ijk}[(V_{KM})_{jl}({\tilde{\nu}}^i_L{\bar{d}}^k_Rd^l_L
+ {\tilde{d}}^l_L{\bar{d}}^k_R {\nu}^i_L +
(\tilde{d} ^k_R
)^* \overline{(\nu ^i_L)^c}d_L^l) - {\tilde{e}}^i_L{\bar{d}}^k_Ru^j_L
- {\tilde{u}}^j_L{\bar{d}}^k_Re^i_L - (\tilde{d} ^k_R
)^* \overline{(e^i_L)^c}u_L^j]
\label{eq:RparityLag}
\end{equation}
where $e$ denotes the electron and ${\tilde{e}}$ it's scalar partner
and similarly for the other particles.

The contributions of the $R$-parity violating interactions to low
energy processes involving no sparticles
in the final state arise
from using the \Rp interactions an even number of times. If two
$\lambda^{\prime}$ 's or $\lambda^{\prime\prime}$ 's with
 different
flavor structure are non-zero, flavor changing low energy processes
can occur. These processes are considered in references
\cite{suzuki} and \cite{barbieri}, respectively.
Therefore, it is usually assumed that either only one $\lambda^{\prime}$
with a
 particular
flavor structure
is non-zero, or that the $R$-parity breaking couplings are diagonal in
generation space. However,
Equation (\ref{eq:RparityLag})
indicates that this does not imply that there is only one set of
interactions with a
particular flavor structure, or even that they are diagonal in flavor
space. In fact, in this case of one
$\lambda^{\prime}_{ijk}\neq 0$, the CKM matrix generates couplings
involving each of the three down-type quarks.
Thus, flavor violation occurs in the down quark sector, though
suppressed by the small values of the off-diagonal CKM elements.
Below, we use these processes to obtain constraints on $R$-parity
breaking, assuming only one
$\lambda^{\prime}_{ijk}\neq 0$.

It would be more
natural to assume that there is only one large \Rp coupling in the
\it weak \rm  basis, i.e., only one
$\bar{\lambda}_{ijk}\neq0$. As we have indicated, this generates many
couplings with different flavor structure in the \it
mass \rm  basis, e.g., many $\lambda_{imn}^{\prime}$'s. It is possible
that
\begin{equation}
\lambda_{imn}^{\prime}\simeq \bar{\lambda}_{ijk}V_{KMjm}\delta_{kn}
\label{assump}
\end{equation}
This will be the case if, for example, the rotation to the mass basis
occurs only for the charge $+2/3$ quark sector.
Then, in addition to the Feynman diagrams that
contribute to the flavor changing neutral current processes when only
one $\lambda_{ijk}^{\prime}$ is present, there are new
contributions
involving the $\lambda_{imn}^{\prime}(m\neq j,n=k)$ vertices. However,
these new
contributions interfere constructively with the operators that are
present in the effective Lagrangian that is
 generated when there is only one non-zero $\lambda_{ijk}^{\prime}$.
So if
these more natural assumptions are made, any constraint found for
$\bar{\lambda}_{ijk}$ is slightly
better than the constraint that is obtained when only one
$\lambda_{ijk}^{\prime}$ is present.

It would seem that the flavor changing neutral current processes may be
rotated away by making a different physical assumption
concerning
which $\not\!\!R_p$ coupling is
non-zero. For example, while leaving the quark fields in the mass
basis, Equation (\ref{physbasis}) gives
\begin{eqnarray}
W_{\not\!R_p}&=&\lambda^{\prime}_{ijk}(N_i(V_{KM})_{jl}D_l - E_iU_j)D^c_k \\
&=&(\lambda^{\prime}_{ijk}V_{KMjl})(N_iD_l - E_i(V_{KMlp}^{-1})U_p)D^c_k \\
\label{twiddle}
&=&\tilde{\lambda}_{ijk}(N_iD_j - E_i(V_{KMjp}^{-1})U_p)D^c_k
\end{eqnarray}
where
\begin{equation}
\tilde{\lambda}_{ijk}\equiv \lambda^{\prime}_{imk}(V_{KM})_{mj}
\end{equation}
With the
assumption that the $\lambda^{\prime}_{ijk}$ coefficients have values such
that only one $\tilde{\lambda}_{ijk}$
is non-zero, there is only one interaction of the form $N_LD_LD^c$.
There is then no longer any flavor violation
in the down quark sector. In particular, there are no \Rp contributions to
the processes discussed below.
But now there are couplings involving each of the three up type quarks. So
these interactions contribute to
FCNC in the up sector; for example, $D^0\!\!-\!\!\bar{D}^0$ mixing. We use
$D^0\!\!-\!\!\bar{D}^0$ mixing to place
constraints on $R$-parity violation assuming only one
$\tilde{\lambda}_{ijk} \neq 0$.
Thus, there is no basis in
which FCNC can be avoided in both sectors.

\subsection{$K^0\!\!-\!\!\bar{K}^0$ Mixing}

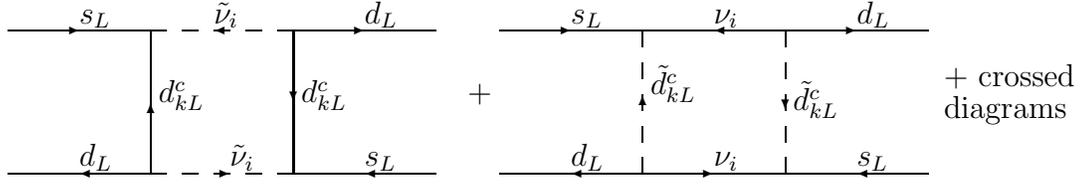
\begin{figure}
\setlength{\unitlength}{0.6pt}
\begin{picture}(610,100)
\put(0,0){
\begin{picture}(300,100)
\put(0,90){\vector(1,0){45}}
\put(45,95){$s_L$}
\put(45,90){\line(1,0){45}}
\put(90,0){\vector(0,1){45}}
\put(95,45){$d_{kL}^c$}
\put(90,90){\line(0,-1){45}}
\put(90,0){\vector(-1,0){45}}
\put(45,+5){$d_L$}
\put(0,0){\line(1,0){45}}
\multiput(90,90)(20,0){2}{\line(1,0){10}}
\put(140,90){\vector(-1,0){10}}
\put(130,95){$\tilde{\nu}_i$}
\multiput(150,90)(20,0){2}{\line(1,0){10}}
\put(180,90){\vector(1,0){45}}
\put(225,95){$d_L$}
\put(225,90){\line(1,0){45}}
\put(180,90){\vector(0,-1){45}}
\put(185,45){$d_{kL}^c$}
\put(180,0){\line(0,1){45}}
\put(270,0){\vector(-1,0){45}}
\put(225,+5){$s_L$}
\put(180,0){\line(1,0){45}}
\multiput(90,0)(20,0){2}{\line(1,0){10}}
\put(130,0){\vector(+1,0){10}}
\put(140,+5){$\tilde{\nu}_i$}
\multiput(150,0)(20,0){2}{\line(1,0){10}}
\put(290,45){$+$}
\end{picture}
}

\put(310,0){
\begin{picture}(330,100)
\put(0,90){\vector(1,0){45}}
\put(45,95){$s_L$}
\put(45,90){\line(1,0){45}}
\put(180,90){\vector(-1,0){45}}
\put(135,95){${\nu}_i$}
\put(90,90){\line(1,0){45}}
\put(90,0){\vector(-1,0){45}}
\put(45,+5){$d_L$}
\put(0,0){\line(1,0){45}}
\multiput(90,90)(0,-20){2}{\line(0,-1){10}}
\put(90,40){\vector(0,1){10}}
\put(95,50){$\tilde{d}_{kL}^c$}
\multiput(90,30)(0,-20){2}{\line(0,-1){10}}
\put(180,90){\vector(1,0){45}}
\put(225,95){$d_L$}
\put(225,90){\line(1,0){45}}
\put(90,0){\vector(1,0){45}}
\put(135,5){${\nu}_i$}
\put(180,0){\line(-1,0){45}}
\put(270,0){\vector(-1,0){45}}
\put(225,+5){$s_L$}
\put(180,0){\line(1,0){45}}
\multiput(180,90)(0,-20){2}{\line(0,-1){10}}
\put(180,50){\vector(0,-1){10}}
\put(185,40){$\tilde{d}_{kL}^c$}
\multiput(180,30)(0,-20){2}{\line(0,-1){10}}
\put(280,55){$+$ crossed}
\put(280,35){diagrams}
\end{picture}
}
\end{picture}
\label{KK}
\caption{$\not\!\!R_p$ contributions to $K^0\!\!-\!\!\bar{K}^0$
mixing with one ${\lambda}^{\prime}_{ijk} \neq 0$. Arrows indicate
flow of propagating left handed fields.}
\end{figure}

With one $\lambda_{ijk}^{\prime} \neq 0$, the interactions of
Equation $(\ref{eq:RparityLag})$
involve down and strange quarks. So, there are contributions to
$K^0\!\!-\!\!\bar{K}^0$
mixing through the box diagrams shown in Figure (1).

Evaluating these diagrams at zero external
momentum and neglecting the down quark masses, the following effective
Hamiltonian is generated
\begin{equation}
{\cal H}^{\Delta
S=2}_{\not\!R_p}=\frac{1}{128{\pi}^2}{\abs{{\lambda}^{\prime}_{ijk}}}^4
\left(\frac{1}{m_{{\tilde{\nu}}_i}^2}+\frac{1}{m_{{\tilde{d}}
_{Rk}}^
2}\right)\big((V_{KM})_{j2}
(V_{KM})_{j1}^*\big)
^2 (\bar{d}_L \gamma^{\mu} s_L)^2
\end{equation}
where $m_{{\tilde{\nu}}_i}$ is the sneutrino mass and
$m_{{\tilde{d}}_{Rk}}$
is the right-handed down squark mass. As this operator
is suppressed by the CKM angles, it is largest when
$\lambda_{ijk}^{\prime}$
is non-zero for $j=1$ or $j=2$.

The SM effective Hamiltonian is \cite{gaillard}
\begin{equation}
{\cal H}^{\Delta S=2}_{SM} =
\frac{G_{F}^2}{4{\pi}^2}{m_c}^2\big((V_{KM})_{12}(V_{KM})_{11}^*\big)^2
  (\bar{d}_L \gamma^{\mu} s_L)^2
\end{equation}
where the CKM suppressed top quark contribution, the up quark mass, and
QCD radiative corrections have been ignored.
As the uncertainty in hadronic matrix elements of the Standard Model
effective
Hamiltonian are at most $40\%$, a conservative
constraint on the \Rp coupling is obtained by demanding that
${\cal L}^{\Delta S=2}_{\not\!R_p} \leq 0.5 {\cal L}^{\Delta
S=2}_{SM}$. This gives the constraint
\begin{equation}
\abs{{\lambda}_{ijk}^{\prime}}\leq 0.08 {\left(\frac{1}{{z_i}^2}
+ \frac{1}{{w_k}^2}\right)}^{-\frac{1}{4}}
\end{equation}
where $z_i=m_{{\tilde{\nu}}_i}/(100\,GeV)$ and
$w_k=m_{{\tilde{d}}_{Rk}}/(100\,GeV)$.
This constraint applies for $j=1$ or $j=2$ and for any $i$ or $k$. The
constraint for $j=3$ is not interesting as the CKM angles suppress the
\Rp operator relative to the
Standard
Model operator.

\subsection{$B^0\!\!-\!\!\bar{B}^0$ Mixing}

The \Rp interactions also contribute to both
$B^0\!\!-\!\!\bar{B}^0$ mixing
and $B_s^0\!\!-\!\!\bar{B}_s^0$
mixing through box diagrams similar to those
given in the previous section. As $B_s^0\!\!-\!\!\bar{B}_s^0$ mixing
is expected
to be nearly maximal, it is not possible at
present to place a constraint on any non-Standard Model effects that would
\it add \rm more mixing. However,
$B^0\!\!-\!\!\bar{B}^0$ mixing has been observed \cite{argus} with a moderate
$x_d\sim 0.7$ \cite{pdg}.
As lattice QCD calculations predict $B_K\sim0.6$ \cite{bernard} and
$B_B\sim1.2$ \cite{martinelli}, it is reasonable to expect
that any $\not \! \! R_p$ contributions
to $B^0\!\!-\!\!\bar{B}^0$ mixing should not exceed $50\%$ of the amount
expected from the Standard Model alone.

The effective Hamiltonian generated by these \Rp processes is
\begin{equation}
{\cal H}_{\not\!R_p}=\frac{1}{128{\pi}^2}{\abs{{\lambda}^{\prime}_{ijk}}}^4
\left(\frac{1}{m_{{\tilde{\nu}}_i}^2}+\frac{1}{m_{{\tilde{d}}_{Rk}}^2}
\right)
\big((V_{KM})_{j3}(V_{KM})_{j1}^*\big)^2
 (\bar{d}_L \gamma^{\mu} b_L)^2
\end{equation}
This is largest when $\lambda_{i3k}^{\prime}$ is non-zero.

The dominant contribution to $B^0\!\!-\!\!\bar{B}^0$ mixing in the
Standard Model is \cite{inami}
\begin{equation}
{\cal H}^{\Delta S=2}_{SM} =
\frac{G_F^2m_t^2}{4{\pi}^2}\big((V_{KM})_{33}(V_{KM})_{31}^*\big)^2G(x_t)
 (\bar{d}_L \gamma^{\mu} b_L)^2
\end{equation}
where $x_t=m_t^2/m_W^2$, and
\begin{equation}
G(x)=\frac{4-11x+x^2}{4(x-1)^2}-\frac{3x^2\ln{x}}{2(1-x)^3}
\end{equation}
For a top mass of $176$ $GeV$, $G(x_t)=0.54$.

This gives the constraint
\begin{equation}
\abs{{\lambda}_{i3k}^{\prime}}\leq 0.77 {\left(\frac{1}{{z_i}^2}
+ \frac{1}{{w_k}^2}\right)}^{-\frac{1}{4}}
\end{equation}
with $z_i$ and $w_k$ as previously defined.

In addition to inducing $B^0\!\!-\!\!\bar{B}^0$ mixing, these
interactions also contribute to the $b\rightarrow s+\gamma$
amplitude.
However, with reasonable values for squark and sneutrino masses,
the constraint is significantly weaker than that found from the
top quark analysis.

\subsection{$K^{+}\rightarrow \pi^{+} \nu \bar{\nu}$}

\begin{figure}
\begin{picture}(300,110)
\put(0,30){\line(1,0){45}}
\put(90,30){\vector(-1,0){45}}
\put(45,35){$s_L$}
\put(90,30){\vector(3,4){27}}
\put(107,71){${\nu}_i$}
\put(117,66){\line(3,4){27}}
\multiput(90,30)(20,0){2}{\line(1,0){10}}
\put(130,30){\vector(1,0){10}}
\put(135,35){$\tilde{d}_k^c$}
\multiput(150,30)(20,0){2}{\line(1,0){10}}
\put(180,30){\line(1,0){45}}
\put(270,30){\vector(-1,0){45}}
\put(225,35){$d_L$}
\put(234,102){\vector(-3,-4){27}}
\put(197,71){${\nu}_i$}
\put(180,30){\line(3,4){27}}
\put(0,0){\vector(1,0){135}}
\put(135,5){$u_L$}
\put(135,0){\line(1,0){135}}
\end{picture}
\label{Kdecay}
\caption{\Rp contribution to $K^{+}\rightarrow \pi^{+} \nu \bar{\nu}$
with one ${\lambda}^{\prime}_{ijk} \neq 0$.}
\end{figure}

The tree level Feynman diagram in Figure (2) generates an effective
Hamiltonian which contributes to the branching
ratio for
$K^{+}\rightarrow \pi^{+} \nu \bar{\nu}$. Using a Fierz rearrangement,
a straightforward evaluation of this diagram
gives
\begin{equation}
{\cal H}_{\not\!R_p}={\frac{1}{2}}
{\frac{{\abs{{\lambda}_{ijk}^{\prime}}}^2}
{m^2_{\tilde{d}_{Rk}}}(V_{KMj1}V_{KMj2}^{\ast})
(\bar{s}_L\gamma^{\mu}d_L)(\bar{\nu}_{Li}\gamma_{\mu}\nu_{Li})}
\end{equation}

There is also a Standard Model contribution to this decay \cite{inami}.
This is an order of magnitude lower than the existing
experimental limit. To obtain a bound on the \Rp coupling, we shall
assume that the \Rp effects dominate the decay
rate.

As the matrix element for this semi-leptonic decay factors into a leptonic
and a hadronic element, the isospin relation
\begin{equation}
\langle \pi^{+}(\bf{p}\rm)|\bar{s}\gamma_{\mu}d|K^{+}(\bf{k}\rm)\rangle
=\sqrt{2}\langle\pi^0(\bf{p}\rm)
|\bar{s}\gamma_{\mu}u|K^{+}(\bf{k}\rm)\rangle
\end{equation}
can be used to relate $\Gamma[K^{+}\rightarrow \pi^{+} \nu \bar{\nu}]$
to $\Gamma[K^{+}\rightarrow \pi^0 {\nu}e^+]$.
The effective Hamiltonian for the neutral pion decay channel
arises from
the spectator decay of the strange quark. It is
\begin{equation}
{\cal H}_{eff}=\frac{4G_F}{\sqrt{2}}V_{KM12}^{\ast}
(\bar{s}_L\gamma^{\mu}u_L)(\bar{\nu}_{Li}\gamma_{\mu}e_{Li})
\end{equation}
So in the limit where the lepton masses can be neglected,
\begin{equation}
\frac{\Gamma[K^{+}\rightarrow \pi^{+} \nu_{i} \bar{\nu}_{i}]}
{\Gamma[K^{+}\rightarrow \pi^0 \nu e^+]}=
{\left(\frac{{\abs{{\lambda}_{ijk}^{\prime}}}^2}
{4G_Fm_{\tilde{d}_{Rk}}^2}
\right)}^2{\left(\frac{\abs{V_{KMj1}V_{KMj2}^{\ast}}}
{\abs{V_{KM12}^{\ast}}}\right)}^2
\end{equation}
This ratio is valid for $i=1,2$ or 3, since in the massless neutrino and
electron approximation, the integrals over phase space in
the numerator and denominator cancel.
So using $BR[K^{+}\rightarrow \pi^{+} \nu \bar{\nu}]\leq5.2\times10^{-9}$
\cite{atiya} ($90\%CL$) and
$BR[K^{+}\rightarrow \pi^0 {\nu}e^+]=0.0482$ \cite{pdg}, the constraint is
\begin{equation}
\abs{\lambda_{ijk}^{\prime}}\leq0.012\left(\frac{m_{\tilde{d}_{Rk}}}
{100\, GeV}\right) (90\%CL)
\end{equation}
for $j=1$ or $j=2$. Using $\abs{V_{KM13}}\geq0.004$ \cite{pdg} and
$\abs{V_{KM23}}\geq0.03$ \cite{pdg}, a conservative upper bound
for $\lambda_{i3k}^{\prime}$ is
\begin{equation}
\abs{\lambda_{i3k}^{\prime}}\leq0.52\left(\frac{m_{\tilde{d}_{Rk}}}
{100\, GeV}\right) (90\%CL)
\end{equation}

\subsection{$D^0\!\!-\!\!\bar{D}^0$ Mixing}

If there is only one $\tilde{\lambda}_{ijk}$ in the \it mass \rm basis,
then from Equation (\ref{twiddle}) it is clear that flavor
changing
neutral current processes will occur in the charge $+2/3$ quark sector.
Rare processes such as $D^0\!\!-\!\!\bar{D}^0$ mixing,
$D^0\rightarrow\mu^+\mu^-$ and $D^+\rightarrow\pi^+l^+l^-$, for example,
may be used to place tight constraints on
$\tilde{\lambda}_{ijk}$. For illustrative purposes, in this section we
will consider $D^0\!\!-\!\!\bar{D}^0$ mixing.

The interactions in Equation (\ref{twiddle}) generate box diagrams
identical to those discussed in the previous
sections if both the internal sneutrino (neutrino) propagators are
replaced with slepton (lepton) propagators and the external
quarks lines are suitably corrected. Using the same approximations
that were made earlier, the \Rp effects generate the
following effective Hamiltonian
\begin{eqnarray}
{\cal H}_{\not\!R_p}&=&\frac{1}{128{\pi}^2}
{\abs{\tilde{\lambda}_{ijk}}}^4
\left(\frac{1}{m_{{\tilde{l}}_i}^2}+\frac{1}{m_{{\tilde{d}}
_{Rk}}^2}\right)\big((V_{KM})_{2j}(V_{KM})_{1j}^*\big)^2
(\bar{c}_L \gamma^{\mu} u_L)^2 \\
&\equiv&G(\tilde{\lambda}_{ijk},m_{{\tilde{l}}_i},m_{{\tilde{d}}_{Rk}})
(\bar{c}_L \gamma^{\mu} u_L)^2
\end{eqnarray}

In the vacuum saturation approximation, the \Rp effects contribute an
amount
\begin{equation}
(\Delta m)_{th}\equiv m_{D_L}-m_{D_S}=\frac{2}{3}f_D^2m_D
ReG(\tilde{\lambda}_{ijk},m_{{\tilde{l}}_i},m_{{\tilde{d}}_{Rk}})
\end{equation}
to the $D_L-D_S$ mass difference. With $f_D=200\, MeV$ \cite{lattice},
$m_D=1864\, MeV$ \cite{pdg}, and
$\abs{(\Delta m)_{exp}}\leq1.32\times10^{-10}\,MeV$ \cite{pdg}($90\%CL$),
the constraint on $\tilde{\lambda}_{ijk}$ for $j=1$ or $j=2$ is
\begin{equation}
\abs{\tilde{\lambda}_{ijk}}\leq 0.16{\left(\left(\frac{100\, GeV}
{{m_{\tilde{l}_i}}}\right)^2 + \left(\frac{100\, GeV}
{{m_{\tilde{d}_{Rk}}}}\right)
^2\right)}
^{-\frac{1}{4}} (90\%CL)
\end{equation}

\section{Top Quark Decay}
\label{sec-top decay}

In the Standard Model, the dominant decay mode for the top quark is
\begin{equation}
t\rightarrow b + W
\end{equation}
with a real $W$ gauge boson produced. This has a partial decay width
\begin{equation}
\Gamma[t\rightarrow W +b]=\frac{G_Fm_t^3}
{8\pi\sqrt2}\abs{V_{tb}}^2(1-x_W^2)(1-2x_W^4+x_W^2)
\end{equation}
where $x_W=m_W/m_t$. The $b$ quark mass has been neglected.

The $R$-parity violating interactions
(see Equation $(\ref{eq:RparityLag})$ with $j=3$)
$\lambda^{\prime}_{i3k}{\tilde{e}}^i_L{\bar{d}}^k_Rt_L$
contribute to the decay
$t_L\rightarrow \tilde{l}_i^+ + d_{Rk}$ at tree level \cite{dreiner},
if kinematically allowed. This is possible only if there
exist sleptons lighter than the top quark.
The partial width for this process is
\begin{equation}
\Gamma[t\rightarrow \tilde{l}_i^+ + d_k]
=\frac{\abs{\lambda_{i3k}^{\prime}}^2 m_t(1-y_i^2)^2}{32\pi}
\end{equation}
with $y_i\equiv m_{\tilde{l_i}}/m_t$ \cite{dreiner}. The mass of the down
type quark has been neglected. If this is the only
non-zero $R$-parity coupling, the two top
quark decay channels are $t\rightarrow b+W $ and $t\rightarrow d_{Rk}+
\tilde{l}_i^+$, with branching
fractions $1-x$ and $x$, respectively.

We assume that the Lightest Supersymmetric Particle (LSP), denoted by
$\tilde{\chi}^0$, is neutral and that the real slepton decays
with 100\% branching fraction to the $\tilde{\chi}^0$ and a lepton.
The presence of a non-zero
$R$-parity breaking coupling implies that the $\tilde{\chi}^0$ is no
longer stable \cite{nilles}. The two dominant decays are
\cite{dreiner} $\tilde{\chi}^0 \rightarrow \nu_i + b +\bar{d_k}$ and
$\tilde{\chi}^0 \rightarrow \bar{\nu_i} + \bar{b} +d_k$.
The LSP decays inside the detector if \cite{dawson}
\begin{equation}
\abs{\lambda_{i3k}^{\prime}} \geq 6\times10^{-5}\sqrt{\gamma}
\left({\left(\frac{100\,GeV}{m_{\tilde{d}_{Rk}}}\right)^2}
+{\left(\frac{100\,GeV}{m_{\tilde{b}_{R}}}\right)}^2 \right)
\left(\frac{100\, GeV}{m_{\tilde{\chi}^0}}\right)^{5/2}
\end{equation}
where $\gamma$ is the Lorentz boost factor of $\tilde{\chi}^0$.
For this decay chain to be kinematically allowed, we require
that $m_{\tilde{\chi}^0} \geq m_b$ for $k=1$ or $k=2$, and
$m_{\tilde{\chi}^0}
\geq 2m_b$ for
$k=3$. Using the
previous equation, the maximum lower bound on $\lambda_{i3k}^{\prime}$
such that the LSP decays
inside the detector is $0.0003\times\sqrt{\gamma}$ for $k=3$, and
$0.002\times\sqrt{\gamma}$ for $k=1$ or $k=2$; all for $300$ $GeV$
squark masses. We shall assume that $\lambda_{i3k}^{\prime}$ is larger
than this value
so that the LSP decays within the detector.

If a top quark decays through this $R$-parity violating process, the
final
state will contain one lepton, at least one $b$ quark and missing
transverse energy. The two novel features of this decay
channel are that it spoils lepton universality and, when $k=3$,
produces a surplus of $b$ quark events.
Both of these signatures can be
used to test the strength of $R$-parity violation.

The CDF collaboration reconstructs $t\bar{t}$ quark events from
observing: (1) dilepton (electron or muon) events
coming from the leptonic decays of both the $W$'s; or (2) one lepton
event arising from leptonic decay of one $W$ and jets from
the hadronic decay of the remaining $W$ boson. CDF also requires
a $b$-tag in the lepton+jets channel.
If the lightest slepton has a mass between 50 and 100 $GeV$, then
the kinematics of the decay $\tilde{l_i}\rightarrow \tilde{\chi}^0
+l_i$ will
be
similar to that of the leptonic decay of the $W$ boson.
A slepton of mass less than 45 $GeV$ is ruled out by the LEP limit on
the $Z$ decay width \cite{pdg}.
If the slepton mass is close to the top mass, then the $b$ quark produced
in the top decay via this channel will
have less energy than the $b$ quark from the top decay via the SM channel.
Also, the lepton from the slepton decay will have more
energy than the lepton from the $W$ decay. These will affect the lepton and
the $b$ quark detection
efficiencies. Although these decay channels will be present for any slepton
lighter than the top
quark, for the purpose of obtaining a constraint, we shall assume that
there
is a slepton with a mass in the range given above.
The presence of the $R$-parity violating coupling will then
contribute signals
to all of these channels.

We assume that the $i=1$ coupling is non-zero. However, all that is
required
is that the
slepton in the generation with the non-zero coupling have a mass in the
range
quoted above, i.e., if $\lambda_{13k}^\prime\neq0$
then we require $50\,GeV<m_{\tilde{e}} <100\,GeV$, and if
$\lambda_{23k}^\prime\neq0$ then we require
$50\,GeV<m_{\tilde{\mu}}<100\,GeV$. Assuming also that the CDF data is
consistent with lepton universality,
the constraints we obtain for $\lambda_{13k}^\prime$ and
$\lambda_{23k}^\prime$ are
identical.

In the $k=1,2$ cases, two $b$ quarks are always produced in a $t\bar{t}$
event. In the $k=3$ case, the LSP decays into
$\bar{b}b{\nu}_i$ or $\bar{b}b\bar{\nu}_i$. Thus, four or six $b$ quarks
may be produced if one or both of the top
quarks decay through
the $R$-parity breaking channel; this possibility must be treated
separately.

\subsection{$\lambda_{i3k}^{\prime}, k\neq 3$}

The branching fraction for the di-electron event is
\begin{equation}
BR[t\bar{t}\rightarrow ee+X]=x^2 +L^2(1-x)^2 +2Lx(1-x)
\end{equation}
with $L=$ leptonic branching fraction of $W$, approximately $1/9$.
The first term arises from both top quarks decaying via the
$R$-parity violating interaction; the second is the Standard Model
contribution; and the third is the contribution from one top quark
decaying through the $R$-parity breaking channel and the other
top quark decaying through the Standard Model channel. The other
branching fractions are
\begin{eqnarray}
BR[t\bar{t}\rightarrow \mu\mu +X]&=&L^2(1-x)^2 \\
BR[t\bar{t}\rightarrow \mu e+X]&=&2(1-x)^2L^2 + 2x(1-x)L \\
BR[t\bar{t}\rightarrow \mu +\hbox{jets}]&=&2(1-x)^2L(1-3L) \\
BR[t\bar{t}\rightarrow e+\hbox{jets}]&=&2(1-x)^2L(1-3L)+2x(1-x)(1-3L)
\end{eqnarray}
The factor of $1-3L$ is the hadronic branching fraction of the $W$ boson.
We have also assumed that the branching fraction for
$\tilde{l} \rightarrow l + \tilde{\chi}^0 $ is close to one.
We are ignoring leptonic events produced from the Standard Model
decay of the $W$ boson into $\tau\nu_{\tau}$.

Two independent constraints on the \Rp interactions may be obtained
from the top quark data.
CDF has observed the $t\bar{t}$
cross section to be $\sigma(t\bar{t})_{exp}=6.8^{+3.6}_{-2.4}$ pb
\cite{CDF1}. The QCD calculation \cite{laenen} gives the value
$\sigma(t\bar{t})_{th}=4.79^{+0.67}_{-0.41}$ pb for $m_t=176\,GeV$.

The first method is to compare the ratio of theoretically
predicted values for the numbers of events found in two channels with
the experimentally observed ratio. For example,
 $\sigma(t\bar{t})_{th}\times BR[t\bar{t}\rightarrow
\mu +\hbox{jets}]\times \int Ldt\times$(detection efficiencies)
is the
number of $\mu$ +jets events that
should have been observed where $\int Ldt$ is the integrated
luminosity. This theoretical prediction contains
uncertainties in both the value for the $t\bar{t}$ production cross
section and in the lepton and the $b$ quark detection efficiences.
In comparing the ratio
\begin{equation}
\left( \sigma(t\bar{t})_{th}\times
BR[t\bar{t}\rightarrow e +\hbox{jets}]\right)
/\left(\sigma(t\bar{t})_{th}\times BR[t\bar{t}\rightarrow \mu
+\hbox{jets}]\right)
\end{equation}
 the uncertainies in the $t\bar{t}$
cross section cancel.
The $b$-detection efficiencies
also cancel. If the electron and the muon
detection efficiences in the lepton + jets channel are equal,
these uncertainties will also cancel. The only
remaining errors are statistical. The CDF collaboration reported
observing
37 $b$-tagged events in the lepton $+\geq 3$ jets
channel. In this set there were 50 $b$-tags, with a background of 22
$b$-tags. A
conservative estimate for the background in the 37 events is 22.
This leaves 15 $t\bar{t}$ events in the lepton +jets
channel. Since no inconsistencies with electron-muon universality have
been reported, a central value of
$7\, \mu$ +jets and $7\, e$ +jets events will be assumed. This
leads to
\begin{equation}
\frac{BR[t\bar{t}\rightarrow e + \hbox{jets}]_{th}}{BR[t\bar{t}
\rightarrow
\mu + \hbox{jets}]_{th}} =
\frac{\#(e + \hbox{jets events})}
{\#(\mu + \hbox{jets events})}
=1^{+a}_{-b}
\label{method1}
\end{equation}
Inserting the theoretical predictions for the branching ratios leads
to the constraint $x<L\,a/(1 +L\,a)$, where $a$ is the
uncertainty in the previous ratio. In this case, $a=b=1/\sqrt{7}$.
This gives $x<0.077$ at $95\% CL$
which leads to
\begin{equation}
\abs{\lambda_{13k}^{\prime}}\leq 0.41\, (95\%\, CL)
\end{equation}
for $k=1$ or $k=2$ and a slepton of mass 100 $GeV$.

A similar analysis may be performed for the dilepton channels.
In principle these channels should lead to a good constraint
since a non-zero $\lambda_{13k}^{\prime}$ coupling will lead to an
excess of electrons observed in the di-electron channel over
the number of
muons observed in the di-muon channel. However at present only a small
number of dilepton events have been observed
and an interesting constraint cannot
be obtained.

In the other method we will compare the number of events produced in a
given channel with the theoretical expectation. The number of produced
events
 is $\sigma [t\bar{t}]_{th}\times BR[t\rightarrow l +
\hbox{jets} ]_{th}\times \int Ldt$. Here
$\sigma [t\bar{t}]_{th}$ is the production cross section calculated in
perturbative QCD for the assumed top quark mass of $176\, GeV$.
We will use the fact that the number of experimentally observed events
in any
given channel is consistent with, within experimental errors, the number
expected in the standard model.
The actual number of events detected depends upon the detection
efficiency. We
will use the number of observed events in any channel to determine the
statistical accuracy with which the rate in that channel is measured,
and then
constrain the strength of the \Rp terms by requiring that the rate is
not changed by more than the error.

This leads
to the constraint
\begin{equation}
\frac{BR[t\bar{t}\rightarrow l + \hbox{jets},x]_{th}}
{BR[t\bar{t}\rightarrow l + \hbox{jets},
x=0]_{th}}=\frac{\sigma[t\bar{t}]
     _{exp}}{\sigma[t\bar{t}]_{th}}
\end{equation}
within theoretical and experimental errors. Using the theoretical
and
experimental values for the production cross sections
\cite{CDF1,laenen} leads
to
\begin{equation}
{\epsilon}^2\leq \frac{BR[t\bar{t}\rightarrow l
+ \hbox{jets},x]_{th}}{BR[t\bar{t}\rightarrow l
+ \hbox{jets},x=0]_{th}}\leq 1+d
\label{method2}
\end{equation}
with $\epsilon=0.9$ and $d=1.37$.
The constraint on $x$ is then
\begin{equation}
x\leq min\left(1-\epsilon,\,\frac{1-2L-\sqrt{(1-2L)
^2-4Ld(1-L)}}{2(1-L)}\right)
\end{equation}
The first entry is the
constraint
from the $\mu$ + jets channel and the second entry is from the
$e$ + jets channel. For
these values of $\epsilon$ and $d$, the constraint is $x\leq
0.1$. For a $100\, GeV$ slepton this translates into the constraint
\begin{equation}
\abs{\lambda_{13k}^{\prime}}\leq 0.46
\end{equation}
for $k=1$ or $k=2$.

\subsection{$\lambda_{i33}^{\prime}$}

For this coupling the analysis of the previous section must be
modified in the lepton + jets channel since the $b$-detection
efficiencies no longer cancel. This is because in the $R$-parity
breaking decay channel three $b$ quarks are produced.
To correct for this, introduce the function $P(k,n)$ that gives
the probability that, given that $n\;b$ quarks are produced,
$k$ of them are detected. Then the number of observed single $b$
quark events expected in the $e +$jets channel is
\begin{eqnarray}
\#(e +\hbox{jets events}) &=& \left(2{(1-x)}^2L(1-3L)P(1,2)
+2x(1-x)(1-3L)P(1,4)\right) \nonumber  \\
                          & &  \times\cal N
\end{eqnarray}
where
\begin{equation}
{\cal N}\equiv\int Ldt \times {\sigma (t\bar{t})}_{th}
\end{equation}
With $P(1,2)\leq P(1,n)$ for $n\geq 2$, then
\begin{equation}
\#(e +\hbox{jets events})\geq \left(2{(1-x)}^2L(1-3L)
+2x(1-x)(1-3L)\right)P(1,2)\times\cal N
\label{inequal}
\end{equation}
These approximations will give a conservative limit for
$\lambda_{133}^{\prime}$.
The analysis of the previous section may now be carried out with
the following restrictions:\\
(i) In comparing the ratio of the numbers of events detected in
two channels with the theoretical prediction, the inequality in
Equation $(\ref{inequal})$ indicates
that only upper limit in Equation $(\ref{method1})$ may used;\\
(ii) In comparing the number of events detected in a channel with
the theoretically predicted value for that channel, only the
upper bound in Equation $(\ref{method2})$ may be used in the $e+$jets
channel, and either limit may be used in the $\mu+$ jets
channel.
With these caveats, a conservative limit on the branching fraction
for $t\rightarrow b+\tilde{l}_i^+$ is then
\begin{equation}
x\leq min \left(L\,a/(1 +L\,a),\,1-\epsilon,\,
\frac{1-2L-\sqrt{(1-2L)^2-4Ld(1-L)}}{2(1-L)}\right)
\end{equation}
For the errors quoted in the previous section, the result is
\begin{equation}
\abs{\lambda_{133}^{\prime}}\leq 0.41\, (95\%\, CL)
\end{equation}
As the $R$-parity breaking decay channels produce three $b$ quarks,
then for moderate values of $\lambda_{133}^{\prime}$ or
$\lambda_{233}^{\prime}$,
semi-leptonic events containing four and six $b$ quarks should
be observable at the Tevatron. The non-observance of these events
should provide the strongest test for the $R$-parity breaking
couplings $\lambda_{133}^{\prime}$ or $\lambda_{233}^{\prime}$.
If limits on the branching fractions for the $t\bar{t}$ pair
to decay into these excess $b$ quark channels are known, then the
$R$-parity branching fraction $x$ is constrained. Namely,
\begin{eqnarray}
&1.& BR[t\bar{t}\rightarrow X+ \geq 3 b's]\leq B_1
\Rightarrow \, x\leq \left(1-\sqrt{1-B_1}\right)\\
&2.&BR[t\bar{t}\rightarrow X+ \geq 3 b's+2e ]\leq B_2
\Rightarrow \, x\leq \frac{\sqrt{L^2+B_2(1-2L)}-L}{1-2L} \\
&3.&BR[t\bar{t}\rightarrow X+ \geq 6 b's +2e ]\leq B_3
\Rightarrow \, x\leq\sqrt{B_3}  \\
&4.&BR[t\bar{t}\rightarrow X+ \geq 3 b's+e ]\leq B_4
\Rightarrow \, x\leq\frac{1}{2}\left(1-\sqrt{1-\frac{2B_4}
{1-3L}}\right) \\
\end{eqnarray}
This constrains $\abs{\lambda_{133}^{\prime}}$.
To constrain $\abs{\lambda_{233}^{\prime}}$,
interchange $e$ with $\mu$ in the previous equations.

The constraints on $\abs{\lambda_{133}^{\prime}}$ and
$\abs{\lambda_{233}^{\prime}}$ found in this section
are comparable to
those obtained from examing
\Rp contributions either to $Z\rightarrow b\bar{b}$ and
$Z\rightarrow l^+l^-$ decays \cite{ellis} or to
forward-backward asymmetry
measurements in $e^+e^-$ collisions \cite{barger}. We have engaged
in this exercise to illustrate how comparable $\not\!\!R
_p$ constraints
may be obtained from analysing top quark decays even though the
experimental and theoretical errors are still large. These
processes will provide much better tests of $R$-parity violation once
more top quark decays are seen.

\section{Summary}

In this paper we have argued that $R$-parity breaking interactions
always lead to flavor changing neutral current processes.
It is possible that there is a single \Rp coupling in the charge
$+2/3$ quark sector.
But requiring consistency with electroweak symmetry breaking
demands that
\Rp couplings involving all the charge $-1/3$ quarks exist. That is,
a single coupling scheme may only be possible in either the
charge $2/3$ or the charge $-1/3$ quark
sector, but not
both. As a result, flavor changing neutral current processes always
exist in one of these sectors.
We have used $K^+ \rightarrow {\pi}^+ \nu \bar{\nu}$,
$K^0\!\!-\bar{K}^0$ mixing, $B^0\!\!-\bar{B}^0$ mixing and
$D^0\!\!-\bar{D}^0$ mixing to constrain the \Rp couplings. The
constraints we obtain
for the first two generations are more stringent than those presently
existing in the literature.

The $R$-parity breaking interactions lead to the top quark decay
$t \rightarrow \tilde{l}_i + d_k$, if the slepton is lighter than
the top quark. Some of the new top quark decays spoil
electron-muon universality or result in $t\bar{t}$ events with more
than
2 $b$ quarks. At present, the CDF collaboration has not reported any
inconsistencies with lepton universality or reported any
 events with
more than 2 $b$ quarks. These decays also lower the branching
fractions for Standard Model top quark decays. These observations
are used to constrain some \Rp couplings.

A list of the known constraints on the $\lambda_{ijk}^{\prime}$
couplings is presented in Table (1). Although several of these
couplings are constrained by different low energy processes,
we have only listed the smallest known upper limit.

\begin{table}
\begin{center}
\begin{tabular}{||l|l||l|l||l|l||}\hline
$\abs{\lambda_{1jk}^{\prime}}$ & &$\abs{\lambda_{2jk}^{\prime}}$
& &$\abs{\lambda_{3jk}^{\prime}}$ &  \\ \hline
111 &0.012$^a$ &211 &0.012$^a$ &311 &0.012$^a$ \\ \hline
112 &0.012$^a$ &212 &0.012$^a$ &312 &0.012$^a$ \\ \hline
113 &0.012$^a$ &213 &0.012$^a$ &313 &0.012$^a$ \\ \hline
121 &0.012$^a$ &221 &0.012$^a$ &321 &0.012$^a$ \\ \hline
122 &0.012$^a$ &222 &0.012$^a$ &322 &0.012$^a$ \\ \hline
123 &0.012$^a$ &223 &0.012$^a$ &323 &0.012$^a$ \\ \hline
131 &0.26$^c$  &231 &0.22$^d$ &331 &0.26$^e$  \\ \hline
132 &0.4$^b$ &232 &0.4$^b$ &332 &0.26$^e$ \\ \hline
133 &0.001$^f$ &233 &0.4$^b$ &333 &0.26$^e$ \\ \hline
\end{tabular}
\end{center}
\label{constraint}
\caption{Constraints on $\abs{\lambda_{ijk}^{\prime}}$ from:(a)
$K^+ \rightarrow \pi ^+ \nu \bar{\nu}$ $(90\% CL)$; (b) top quark
decay $(95\% CL)$; (c) atomic parity violation and $eD$ asymmetry
$(90\% CL)$ \protect\cite{barger}; (d) $\nu_{\mu}$ deep-inelastic
scattering $(95\% CL)$ \protect\cite{barger}; (e) partial $Z^0$
decay width $(95\% CL)$ \protect\cite{ellis}; (f) $\nu_e$ mass
$(90\% CL)$ \protect\cite{godbole}. All limits are for $100\,
GeV$ sparticle masses.  }
\end{table}

The tightest constraint is on $\abs{\lambda_{ijk}^{\prime}}$ for
$j=1,2$ and any $i$ and $k$. This comes from the rare decay
$K^+ \rightarrow \pi ^+\nu \bar{\nu}$. With the exception of
$\lambda_{133}^{\prime}$, the constraints on the third quark
generation couplings are only of order $e/\sin \theta _w$.
Once more top quark decays are
 observed the signatures discussed in this paper will more
tightly constrain these couplings.

\section{Acknowledgements}
The authors would like to acknowledge both I. Hinchliffe and
M. Suzuki for many useful discussions.
This work was supported in part by the Director, Office of
Energy Research, Office of High Energy and Nuclear Physics,
Division of
High Energy Physics of the U.S. Department of Energy under Contract
DE-AC03-76SF00098 and in part by the National Science Foundation under
grant PHY-90-21139.
M.G. would also like to thank the Natural Sciences and Engineering
Research Council of Canada (NSERC) for their support.


\begin{thebibliography}{99}
\bibitem{nilles}For reviews of supersymmetry and
supersymmetry phenomenology, see: \\
P. Fayet and S. Ferrara, \prpt{5}{249}{77};
H.P. Nilles, \prpt{110}{1}{84}; M.F. Sohnius, \prpt{128}{2}{85};
I. Hinchliffe, \arn{36}{505}{86}.
\bibitem{suzuki}L.J. Hall and M. Suzuki, \np{B231}{419}{84}.
\bibitem{carlson} C.E. Carlson, P. Roy and M. Sher, \pl{B357}{99}{95}.
\bibitem{hall}S. Dimopoulos and L.J. Hall, \pl{B207}{210}{87}.
\bibitem{barger}V. Barger, G.F. Giudice and T. Han, \pr{D40}{2987}{89}.
\bibitem{godbole}R. Godbole, P. Roy and X. Tata, \np{B401}{67}{93}.
\bibitem{dawson}S. Dawson, \np{B261}{297}{85}.
\bibitem{choudhury} G. Bhattacharyya and
D. Choudhury, \mpl{A10}{1699}{95}.
\bibitem{ellis} G. Bhattacharyya, J. Ellis and
K. Sridhar, \mpl{A10}{1583}{95}.
\bibitem{mohapatra} R. Mohapatra, \pr{D34}{3457}{86}.
\bibitem{D0}S. Abachi \etal, \prl{74}{2632}{95}.
\bibitem{CDF1}F. Abe \etal, \prl{74}{2626}{95}.
\bibitem{pdg}\it Review of Particle Properties\rm, \pr{D50}{1177}{94}.
\bibitem{ma} E. Ma and P. Roy, \pr{D41}{988}{90};
E. Ma and D. Ng, \pr{D41}{1005}{90}.
\bibitem{kobayashi}N. Cabibbo, \prl{10}{531}{63};
M. Kobayashi and T. Maskawa, \prgth{49}{652}{73}.
\bibitem{barbieri} R. Barbieri and A. Masiero, \np{B267}{679}{86}.
\bibitem{gaillard}M.K. Gaillard and B.W. Lee, \pr{D10}{897}{74}.
\bibitem{argus}C. Albajar \etal, \pl{B186}{247}{87};
H. Albrecht \etal, \pl{B197}{452}{87}.
\bibitem{bernard} C. Bernard, \npp{B34}{47}{94};
S. Sharpe, Lectures given at the Theoretical Advanced
Study Institute in Particle
Physics (TASI 94), Boulder, CO, 29 May - 24 Jun 1994.
\bibitem{martinelli}G. Martinelli, in Proceedings of the
6th Rencontres De Bois, Bois, France 20 - 25 Jun 1994.
\bibitem{inami}T. Inami and C.S. Lim, \prgth{65}{292}{81}.
\bibitem{atiya}M.S. Atiya \etal, \prl{70}{2521}{93};
(erratum) M.S. Atiya \etal, \prl{71}{305}{93}.
\bibitem{lattice}J. Shigemitsu, in \it Proceedings of the
XXVII International Conference on High Energy
Physics \rm (Glasgow, Scotland, UK, July 1994), edited by
P.J. Bussey and I.G. Knowles (Institute of Physics Publishing,
Bristol and Philadelphia, 1995).
\bibitem{dreiner}H. Dreiner and R.J.N. Phillips, \np{B367}{591}{91}.
\bibitem{laenen}E. Laenen, J. Smith and
W.L. van Neerven,\pl{B321}{254}{94}.
\end{thebibliography}
\end{document}